# QUANTUM HALL EFFECT ON DIRAC ELECTRONS IN MODULATED GRAPHENE

**MUHAMMAD ARSALAN ALI**

(**150938**)

**BS PHYSICS (FALL-2015)**

**THESIS SUPERVISOR: DR. RUBINA NASIR**

ASSISTANT PROFESSOR

DEPARTMENT OF PHYSICS

AIR UNIVERSITY

PAF COMPLEX E-9, ISLAMABAD

**Abstract**

Theoretical investigation of Dirac electrons in electrically modulated graphene under perpendicular magnetic field B is presented. We have carried out a detailed study of modulation effect on Dirac electrons, which determine its electrical transport properties. The periodic potential broadens the Landau levels (LL), which oscillate with magnetic field B and a comparison made with two-dimensional electron gas system (2DEGS). We have found the effect of Hall conductivity on electronic conduction in this system. In addition, we find that Hall conductivity exhibits Weiss oscillations and Shubnikov de Haas (SdH) oscillations. The effect of temperature and the period of periodic potential on these oscillations are studied in this work. Furthermore, an integral quantum Hall effect in graphene is also discussed.

# Table of Contents

# Chapter One
# Introduction

## 1.1 Graphene

Graphene is a two-dimensional material, in which carbon atom is bonded in a honeycomb lattices through $Sp^2$ hybridization bonding. The nature of quasiparticles in graphene is very different from the two-dimensional electron gas system.

Graphene has been a puzzling element for scientist for many decades. Graphene was first discussed by Philip Wallace in 1946 when he was studying the graphite. If sheets of graphene stacked on top of each other, you get graphite. He used theoretical data for graphene to extract facts about the different properties of graphite. Wallace believed that his theoretical effort towards graphene would help in its construction in the near future [1]. Theoretical researched for graphene boosted when fullerenes discovered in 1980.

In 2004 Andre Geim and Konstantin Novoselov first prepared Wallace's graphene by applying a piece of graphite on sticky tape, to pull off flakes of graphite. The procedure repeated again and again to get a layer of graphite flakes of atomic size (one atom thick). This technique to prepare graphene is called the scotch tape method [2]. Geim and Novoselov awarded Noble price in 2010.

Geim and Novoselov discovery of graphene in the laboratory was incredible, for the scientist to think how the physics laws modified if electron motion is restricted to two dimensions instead of three. What will be the effective mass of electron in graphene system? How the energy dispersion relation will take form for two-dimensional graphene system? Is graphene follow the mathematics derive for two-dimensional electron gas (2DEGS)?

Graphene has incredible mechanical properties like stiffness, strength and toughness that make graphene unique from other material. Graphene is the strongest material ever tested with breaking strength $42\,\text{Nm}^{-1}$ [3]. Graphene also have high value of mobility and thermal conductivity. At room temperature the electron mobility in graphene system is $15000\,\text{cm}^{-1}\text{V}^{-1}\text{s}^{-1}$ [4].

The nature of quasiparticles in graphene is very different from the two-dimensional electron gas system. When we study the band structure of graphene and we have found that the electrons and hole band touch in two pints in Brillouin zone. These points are called Dirac points because at these points, the fermions obey the massless Dirac equation and lead to linear dispersion relation $E = \hbar v_F\, k$ (where $v_F \cong 10^6 \text{ms}^{-1}$).

Due to this relativistic behavior of fermions in graphene and its unconventional Landau levels, leads to the anomalous quantum Hall effect [5]. The conductivity and resistivity properties of graphene and the effect of periodic potential on Dirac fermions, in this system, has been a topic of great interest. The electric modulation in this system is carried out by depositing an array of parallel metallic strips on the surface [6].In this work, we discussed the motion of conduction electrons under the combined influence of a magnetic field and a weak electric modulation in this system. The electronic transport in this system is highly affected by electric modulation. In a result of electric modulation Commensurability (Weiss) oscillations and Shubnikov de Haas (SDH) oscillations are prevailed. As long the magnetic field is small these oscillations are periodic in 1/B. To make a complete study of the effect of weak electric modulation on electronic transport, we determine the Hall conductivity. In the end, we discussed the quantum Hall effect on Dirac electron in electrically modulated graphene [7].

**1.2 Motivation**

The electronic transport in graphene system is very much different from other two-dimensional materials and semiconductor used in electronic industry. The discovery of graphene changed mindset for engineer and scientist to make electronic device from graphene due to its properties like high mobility and thermal conductivity. At room temperature the electron mobility in graphene system is 15000 $\text{cm}^{-1}\text{V}^{-1}\text{s}^{-1}$ [4].

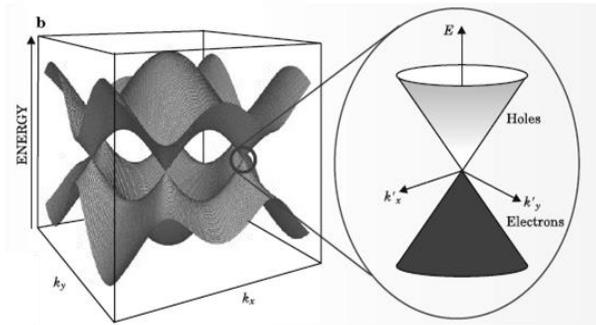

Figure1.1: Electronic Band Structure of Graphene.

Graphene has incredible mechanical Properties like stiffness, strength and toughness that make graphene unique from other material. Graphene is the strongest material ever tested with breaking strength 42 Nm$^{-1}$ [3]. The nature of quasiparticles in graphene is very different from the two-dimensional electron gas system. When we study the band structure of graphene and we have found that the electrons and hole band touch in two pints in Brillouin zone. These points are called Dirac points because at these points, the fermions obey the massless Dirac equation and lead to linear dispersion relation $E = \hbar v_F k \ (where \ v_F \cong 10^6 ms^{-1})$.

In the presence of perpendicular magnetic field B, the electronic transport in graphene system show different behavior as compare to two-dimensional electron gas system.

The one-electron Hamiltonian for graphene under perpendicular magnetic field is given as:

$$H = v_f \sigma \cdot (p + eA) \quad (1.1)$$

The corresponding Landau level energy eigenvalue without modulation are given as:

$$E_n = \hbar \omega_g \sqrt{n} \quad (1.2)$$

Where n is integer and $\omega_g = v_f \sqrt{2eB/\hbar}$.

The energy spectrum for electrons in 2DEGs is given as:

$$E_n = \hbar \omega_c \left( n + \frac{1}{2} \right), \quad (1.3)$$

where $\omega_c = \frac{eB}{mc}$ is the cyclotron frequency.

From Equation (1.01) and (1.02) we can see that the energy spectrums, the Landau level energy spectrum for unmodulated graphene is significantly different from conventional 2DEGS.

The mobility and thermal conductivity of graphene and the effect of periodic potential on Dirac fermions, in this system, has been a topic of great interest. The electronic transport in this system is highly affected by electric modulation. In a result of electric modulation Commensurability (Weiss) oscillations and Shubnikov de Haas (SdH) oscillations are prevailed. Due to this relativistic behavior of fermions in graphene and its unconventional Landau levels, leads to the anomalous quantum Hall effect [5].

## 1.2 Thesis Overview

This report is divided into four chapters. In the first chapter the introduction is presented.

In the second chapter Literature review is presented, we study quantum Hall effect, two-dimensional electron gas system and effect of Shubnikov de Haas (SDH) oscillations.

In the third chapter we study the effect of weak electric modulation on the electronic transport of graphene monolayer system, under a perpendicular magnetic field B is applied on this system. We determine the Landau energy Eigen values using Perturbation theory after applying periodic potential as a modulation. Electric modulation introduces a length scale and period of modulation, which will affect the electronic transport of graphene system. In a result of electric modulation Commensurability (Weiss) oscillations and Shubnikov de Haas (SDH) oscillations are prevailed. We will also discuss the effect of modulation on Landau energy spectrum of graphene and compare it with the energy spectrum of two-dimensional electron gas system.

In the four chapter, we determine the expression for Hall conductivity for graphene monolayer under perpendicular magnetic field B. Using this expression we plot the graph of conductivity and resistivity at different temperature (T=2K and T=6K) and study the effect of temperature on Commensurability (Weiss) oscillations and Shubnikov de Haas (SDH) oscillations. At the end we study the quantum Hall effect in graphene system in detail. In the end the results and discussion are presented.

# Chapter Two
# Literature Review

## 2.1 Classical Hall Effect

In 1879 Edwin Hall discovered the classical Hall Effect. When the electrons motion is restricted to move in two-dimensional plane and strong magnetic field is applied the most wonderful and surprising results is observed which Hall Effect is. Before discussing the quantum Hall effect we take a quick review of classical Hall Effect.

Systematic diagram for classical effect is as shown in figure (2.1). The electron motion are restricted to xy-plane and current $I_x$ is passing through the sample in the x-direction. A perpendicular magnetic field B is applied in the z-direction. So a voltage is produced in the y-direction which is known as Hall Voltage ($V_H$).

## 2.2 Classical Motion

The perpendicular magnetic field B restrict the particle to follow circular motion in the xy-plane and an electric field E is applied which will accelerated the charges and cause a current in the x-direction. The equation motion become:

$$m\frac{d\mathbf{v}}{dt} = -e\mathbf{E} - e(\mathbf{v} \times \mathbf{B}) \tag{2.1}$$

The current density is related to velocity as:

$$\mathbf{J} = -ne\mathbf{v} \tag{2.1a}$$

And

$$\mathbf{J} = \sigma \mathbf{E} \tag{2.1b}$$

$$\begin{bmatrix} J_x \\ J_y \end{bmatrix} = \begin{bmatrix} \sigma_{xx} & \sigma_{xy} \\ \sigma_{yx} & \sigma_{yy} \end{bmatrix} \begin{bmatrix} E_x \\ E_y \end{bmatrix} \tag{2.2}$$

Where n is the number of charge carriers and v is the velocity of charge carries.

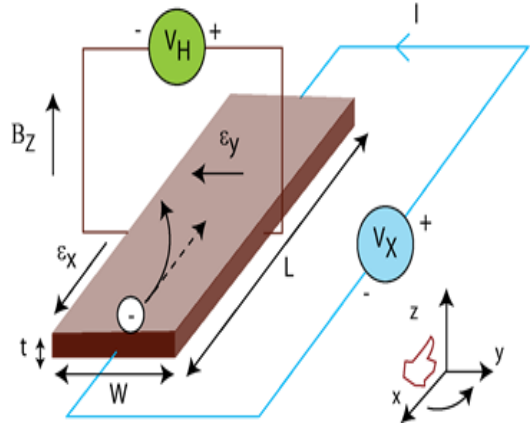

Figure 2.1: The classical Hall Effect.

$$\frac{e\mathbf{E}}{m} = \frac{e}{m}(\mathbf{v} \times \mathbf{B}) \qquad (2.3)$$

By writing in matrix notation and using equation

$$\begin{bmatrix} 1 & \omega_g \\ -\omega_g & 1 \end{bmatrix} \begin{bmatrix} J_x \\ J_y \end{bmatrix} = \frac{ne^2}{m} \begin{bmatrix} E_x \\ E_y \end{bmatrix}$$

Where $\omega_g = \frac{eB}{m}$ is the cyclotron frequency of electron

$$\begin{bmatrix} J_x \\ J_y \end{bmatrix} = \frac{ne^2}{m} \begin{bmatrix} 1 & 1/\omega_g \\ -1/\omega_g & 1 \end{bmatrix} \begin{bmatrix} E_x \\ E_y \end{bmatrix} \qquad (2.4)$$

$$\begin{bmatrix} J_x \\ J_y \end{bmatrix} = \begin{bmatrix} \frac{ne^2}{m} & ne/B \\ -ne/B & \frac{ne^2}{m} \end{bmatrix} \begin{bmatrix} E_x \\ E_y \end{bmatrix} \qquad (2.5)$$

As $\sigma_{xx} = \sigma_{yy} \cong \frac{ne^2}{m}$ and $\sigma_{xy} = -\sigma_{yx} \cong \frac{ne}{B}$, so Equation (2.5) become

$$\begin{bmatrix} J_x \\ J_y \end{bmatrix} = \begin{bmatrix} \sigma_{xx} & \sigma_{xy} \\ \sigma_{yx} & \sigma_{yy} \end{bmatrix} \begin{bmatrix} E_x \\ E_y \end{bmatrix} \qquad (2.6)$$

Hence

$$\sigma = \begin{pmatrix} \sigma_{xx} & \sigma_{xy} \\ \sigma_{yx} & \sigma_{yy} \end{pmatrix} \qquad (2.7a)$$

This is called conductivity tenors. The off diagonal element is responsible for Hall conductivity. Back to classical Hall effect, an electric field E is applied which will accelerated the charges in the x-direction and cause a current density $J_x$. Due to presence of perpendicular magnetic field in the z-direction this current bends towards the y-direction and an electric field $E_y$ start to build in the y-direction until electric field $E_y$ become equal to magnetic field B. Due to electric field $E_y$ a voltage is induced, which known as Hall voltage and given as:

$V_{hall} = E_y w$   (Where w is the width of sample)

## 2.3 Quantum Hall Effect

In 1980 von Klitzing discovered the quantum nature of Hall Effect. Quantum-mechanically when two-dimensional electron systems is subjected to low temperatures and strong magnetic fields, the Hall conductivity $\sigma$ undergoes quantum Hall transitions by taking only the quantized values and series of step called plateau, appear as a function of magnetic field. These plateaus can be understood in better way with help of Landau level concept. When a magnetic field is applied the available splits into different energy levels relative to Fermi energy called Landau levels, separated by the cyclotron energy. The region between two Landau levels is consider as forbidden states. When modulation is applied the Landau levels further splits.

The classical Hall resistance is given as $B/ne$ and the energy of each landau level is $eB/\hbar$. At these plateaus the resistivity takes the value

$$\rho_{xy} = \frac{1}{v}\left(\frac{\hbar}{e^2}\right) \tag{2.8}$$

Where *v* is known as the filling factor, and can take on integer (*v* = 1, 2, 3,…)
The difference between classical Hall Effect and quantum Hall effect is that the Hall resistance can take the quantized value.

# Chapter Three

# Electrically Modulated Graphene

## 3.1 Formulation

We suppose that a monolayer graphene system in xy-plane and perpendicular magnetic field B is applied in z-directions as shown in figure (1-1). The one-dimensional weak periodic potential U(x) in the x direction is applied as a modulation.

The one electron Hamiltonian is given as:

$$H = v_f \boldsymbol{\sigma} \cdot (\boldsymbol{p} + e\boldsymbol{A}) \tag{3.1}$$

Where, $\boldsymbol{\sigma} = \{\sigma_x, \sigma_y\}$ is Pauli matrices, p is the momentum, $v_f \cong 10^6 ms^{-1}$ is the electron velocity in graphene and A is the vector potential. For unperturbed system U(x) will be zero and in Landau gauge vector potential $A = (0, B_x, 0)$, the normalized Eigen function of Equation (3.01) are given by:

$$\Psi_{n,k_y} = \frac{e^{ik_y y}}{\sqrt{2l_y l}} \begin{pmatrix} -i\phi_n\left(\frac{x+x_o}{l}\right) \\ \phi_{n-1}\left(\frac{x+x_o}{l}\right) \end{pmatrix} \tag{3.2}$$

Where $\phi_n = \frac{1}{\sqrt{2^n n!\sqrt{\pi}}} e^{-\frac{x^2}{2}} H_n(x)$ and $\phi_{n-1} = \frac{1}{\sqrt{2^{n-1}(n-1)!\sqrt{\pi}}} e^{-\frac{x^2}{2}} H_{n-1}(x)$ are harmonic oscillator wave functions, n is the Landau level index, $L_y$ is length of two-dimensional graphene in the y direction and $l = \sqrt{\frac{\hbar}{eB}}$ is the magnetic length. The eigenvalue corresponding to equation (3.1) is read as:

$$E_n = \hbar \omega_g \sqrt{n} \tag{3.3}$$

Where $\omega_g = v_f \sqrt{2eB/\hbar}$ is the cyclotron frequency of Electrons in graphene.

After applying electric modulation, the one-electron Hamiltonian for graphene is given as:

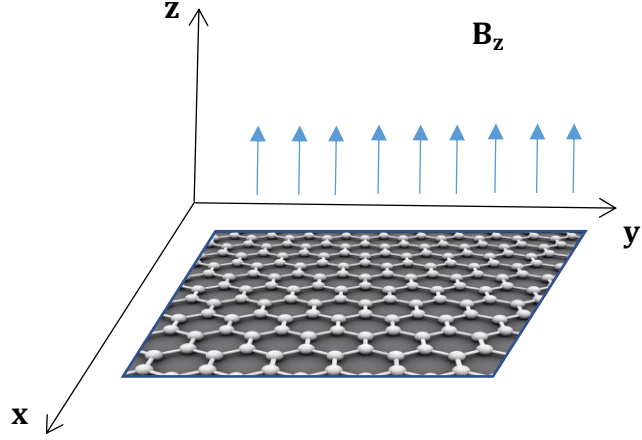

Figure 3.1: Electrically modulated Graphene.

Figure 3.2(a): Landau bandwidth in monolayer graphene as a function of magnetic field: solid curve is the exact result of bandwidth and dash curve is asymptotic result of bandwidth.

Figure 3.3: Electrically modulated Graphene.

We can approximate the periodic potential $U(x) = V_o cosKx$, where $K = \frac{2\pi}{a}$ with $a$ is the period of modulation and $V_o$ is the amplitude of electric modulation.

As the applied modulation is weak, so we can use perturbation theory to calculate energy eigenvalue of weak electrically modulated graphene.

Consider $H_p$ is the Hamiltonian after applied the modulation U(x) and $H_o$ is the Hamiltonian of unmodulated graphene under perpendicular magnetic field B. Then the total Hamiltonian read as:

$$H = H_p + H_o \qquad (3.5)$$

$$H|\psi> = (H_p + H_o)|\psi> = E_n|\psi> \qquad (3.6)$$

The first order perturbation theory is given as:

$$E_n = E_n^{(0)} + \langle \varphi_n | H_p | \varphi_n \rangle \qquad (3.7)$$

Using First order perturbation theory

$$E_{n,k_y} = E_n + \Delta E_{n,k_y} \tag{3.8}$$

Now we calculate energy Eigen value for perturbed system

$$\Delta E_{n,k_y} = \langle \varphi_n | H_p | \varphi_n \rangle$$

$$\Delta E_{n,k_y} = \int_{-\infty}^{+\infty} \int_0^{L_y} \Psi_{n,k_y}^* \, U(x) \, \Psi_{n.k_y} dx dy \tag{3.9}$$

Putting the values in equation (3.9)

$$\Delta E_{n,k_y} = \frac{V_o}{2L_y l} \int_{-\infty}^{+\infty} \int_0^{L_y} \left\{ \phi_n^*\left(\frac{x+x_0}{l}\right) \phi_n\left(\frac{x+x_0}{l}\right) + \phi_{n-1}^*\left(\frac{x+x_0}{l}\right) \phi_{n-1}\left(\frac{x+x_0}{l}\right) \right\} \cos K(x) dx dy \tag{3.10}$$

By making substitution

$$x' = \frac{x+x_0}{l} \qquad\qquad dx = l dx'$$

After substitution, $\phi_n$ and $\phi_{n-1}$ are harmonic oscillator wave functions becomes:

$$\phi_n(x') = \frac{1}{\sqrt{\sqrt{\pi}\, 2^n n!}} H_n(x') e^{-\frac{x'^2}{2}} \tag{3.11}$$

$$\phi_{n-1}(x') = \frac{1}{\sqrt{\sqrt{\pi}\, 2^{n-1}(n-1)!}} H_{n-1}(x') e^{-\frac{x'^2}{2}} \tag{3.12}$$

After defining the new variable we solve the equation (3.10)

$$\Delta E_{n,k_y} = \frac{V_o l}{2L_y l} \int_0^{L_y} \int_{-\infty}^{+\infty} \frac{1}{\sqrt{\pi}\, 2^n n!} \left\{ \begin{matrix} H_n^2(x') e^{-x'^2} \\ cosK(x'l - x_0) \end{matrix} \right\} \quad (3.13)$$

$$+ \frac{1}{\sqrt{\pi}\, 2^{n-1}(n-1)!} \left\{ \begin{matrix} H_{n-1}^2(x') e^{-x'^2} \\ cosK(x'l - x_0) \end{matrix} \right\} dy\, dx'$$

Using Formaula $\cos(\alpha + \beta) = cos\alpha\, cos\beta + sin\alpha\, sin\beta$

$$= \frac{V_o}{2L_y} \int_0^{L_y} \int_{-\infty}^{+\infty} \left[ \frac{1}{\sqrt{\pi}\, 2^n n!} H_n^2(x') e^{-x'^2} \{cosKx'l\, cosKx_o + sinKx'l\, sinKx_o\} \right. \quad (3.14)$$

$$\left. + \frac{1}{\sqrt{\pi}\, 2^{n-1}(n-1)!} H_{n-1}^2(x') e^{-x'^2} \{cosKx'\, lcosKx_o + sinKx'\, lsinKx_o\} \right] dy\, dx'$$

Using orthogonality condition, the sin term vanish, so above equation takes form:

$$\Delta E_{n,k_y} = \frac{V_o}{2L_y} \int_0^{L_y} \int_{-\infty}^{+\infty} \left[ e^{-x'^2} cosKx'l\, cosKx_o \left\{ \frac{1}{\sqrt{\pi}\, 2^n n!} H_n^2(x') + \frac{1}{\sqrt{\pi}\, 2^{n-1}(n-1)!} H_{n-1}^2(x') \right\} \right] dy\, dx' \quad (3.16)$$

$$\Delta E_{n,k_y} = V_o cosKx_o \frac{1}{2L_y} \quad (3.15)$$

$$\times L_y \int_{-\infty}^{+\infty} \left[ e^{-x'^2} cosKx'l \left\{ \frac{1}{\sqrt{\pi}\, 2^n n!} H_n^2(x') + \frac{1}{\sqrt{\pi}\, 2^{n-1}(n-1)!} H_{n-1}^2(x') \right\} \right] dx'$$

The integral appear in equation (3.16) can be solved by using standard integral:

$$\Delta E_{n,k_y} = V_o cosKx_o \begin{bmatrix} \sqrt{\pi}\, 2^n n!\, e^{-\frac{u^2}{2}} L_n(u) \left\{ \frac{1}{\sqrt{\pi}\, 2^n 2^{-1} n!} \right\} + \\ \sqrt{\pi}\, 2^n 2^{-2}(n-1)!\, e^{-u^2/2} L_{n-1}(u) \left\{ \frac{1}{\sqrt{\pi}\, 2^n 2^{-1}(n-1)!} \right\} \end{bmatrix}$$

$$\int_0^\infty e^{-y^2}[H_n(y)]^2 \cos(\sqrt{2}\beta y)\, dy = \sqrt{\pi}\, 2^{n-1} n!\, e^{-\frac{\beta^2}{4}} L_n(\beta^2)$$

$$\beta^2 = u^2 = \left(\frac{Kl}{2}\right)^2$$

Further solving we arrive at the following:

$$\Delta E_{n,k_y} = V_o e^{-u^2/2} \cos Kx_o \left[\frac{L_n(u) + L_{n-1}(u)}{2}\right] \tag{3.17}$$

Putting equation (3.17) in equation (3.8) and arrive at following Expression:

$$E_{n,k_y} = E_n + V_{n,B} \cos Kx_o \tag{3.18}$$

Where $V_{n,B} = V_o e^{-u^2/2} \left[\frac{L_n(u)+L_{n-1}(u)}{2}\right]$ is bandwidth of landau Levels, $L_n(u)\ and\ L_{n-1}(u)$ are Laguerre polynomials.

## 3.2 Asymptotic expression of for Landau bandwidth

The asymptotic expression of for Landau bandwidth Can be obtained by using the following asymptotic expression as n >>1

$$e^{-\frac{u}{2}} L_n(u) = \frac{1}{\sqrt{\pi\sqrt{nu}}} \cos\left(2\sqrt{nu} - \frac{\pi}{4}\right) \tag{3.19}$$

Using above expression, the half bandwidth $V_{n,B}$ becomes:

$$V_{n,B} = V_o \sqrt{\frac{2}{\pi K R_c}} \cos\left(KR_c - \frac{\pi}{4}\right) \tag{3.20}$$

Where $R_c = k_F l$ is the classical cyclotron orbit, $k_F = \sqrt{2\pi n_e}$ and $n_e$ is the electron number density. The above expression is also function of magnetic field B because $R_c = k_F l = k_F \sqrt{\hbar/eB}$.

There are following two condition:

For maxima of Landau bandwidth:

$$\frac{2R_c}{a} = i + \frac{1}{4} \qquad (3.21)$$

For minima of Landau bandwidth:

$$\frac{2R_c}{a} = i - \frac{1}{4} \qquad (3.22)$$

Where $i = |1, 2, 3, 4, \ldots|$

Equation (3.15) is called board band condition and Equation (3.17) is called flat band condition for Landau Level. The half Landau bandwidth of electrically modulated graphene as shown in Figure (3.2a). There are two curves in Figure (3.2a) solid curve and dash curve: The Solid curve is the extract result of half bandwidth; $V_{n,B} = V_o e^{-u^2/2} \left[\frac{L_n(u) + L_{n-1}(u)}{2}\right]$. The dash curve is the asymptotic expression of Landau band width.

From the figure (3.2a) it is clear that oscillations of the Landau bandwidth are origin of Weiss oscillations. From the figure (3.2a) it is also clear that at low magnetic field the Weiss oscillations appear expect the region of very strong magnetic field where SdH oscillations become dominant because the exact result and its asymptotic expression are same at low magnetic field but a jump in the solid curve appears as the value magnetic field is increased. The jump in the solid curve is due to SdH oscillation.

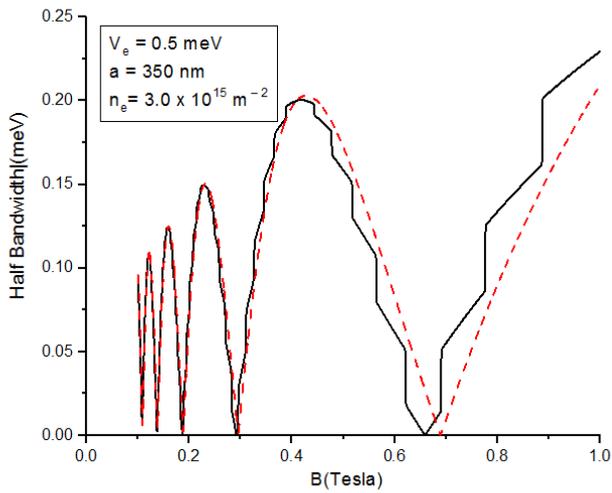

Figure 3.4(a): Landau bandwidth in monolayer graphene as a function of magnetic field: solid curve is the exact result of bandwidth and dash curve is asymptotic result of bandwidth.

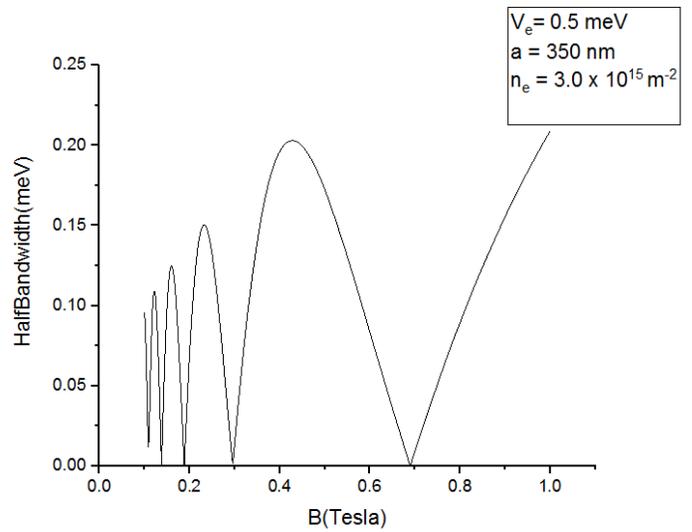

Figure 3.7(b): The Bandwidth of the Landau Level in monolayer graphene as a function of magnetic field.

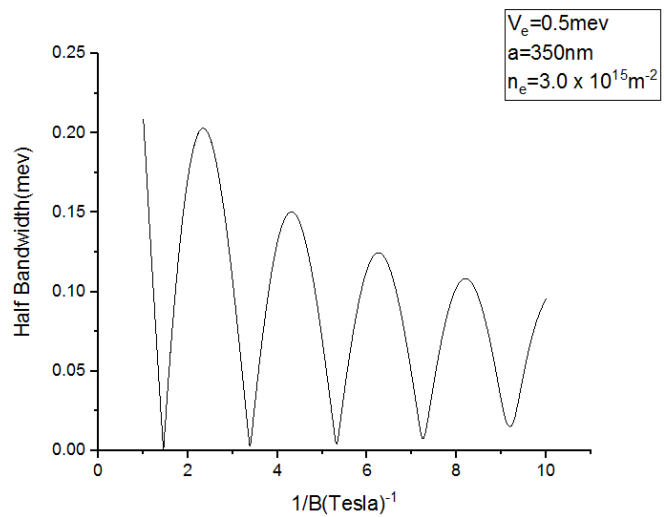

Figure 3.2(c): The Half Band width of Landau Level for monolayer graphene as a function of 1/B.

## 3.3 Energy dispersion relation for electrically modulated graphene

Equation (3.14) shows the broadening of the energy levels of graphene in non-uniform under perpendicular magnetic field B and a weak electric modulation. The broadening of Landau levels is non-uniform because of $E_n$ Proportional to the square root of Landau index n. Also, Equation (3.14) shows that Landau bandwidth oscillates as Laguerre Polynomials are function of Landau level index n. From Figure (3.2b) it is proved that the Landau bandwidth broadening is non-uniform and have oscillating behavior.

Now we start a comparison between 2DEGS and two-dimensional graphene system. From the Figure (3.2b) it is clear that Landau Level spectrum is very much different from conventional 2DEGS. The first term in the equation (3.14), which gives Landau energy spectrum is $E_n = \hbar \omega_g \sqrt{n}$ and $\omega_g = v_f \sqrt{2eB/\hbar}$ while for 2DEGS the expression for energy spectrum is $E_n = \hbar \omega_g (n + \frac{1}{2})$ and $\omega_c = eB/m$.

From the above discussion we find two basis difference between 2DEGS and two-dimensional graphene system: First, In Landau half bandwidth we have an average of two successive Laguerre term while in 2DEGS it has only one Laguerre term. Second, the Landau energy spectrum has a square root of Landau index n which is not present in 2DEGS.

Figure 3. 2c shows the half bandwidth of Landau Level for monolayer graphene as a function of 1/B. From the Literature review we seen that SdH oscillations are periodic with 1/B. So Figure 3.2c confirm the presence of SdH oscillations.

# Chapter Four

# Hall Conductivity

## 4.1 Conductivity tensor and standard expression for resistivity tensor $\rho_{\mu\nu}$

We consider a two-dimensional monolayer graphene system in the xy-plane and perpendicular magnetic field B is applied in z-directions. The one-dimensional weak periodic potential U(x) in the x direction is applied as a modulation. We can approximate the periodic potential as $(x) = V_o \cos Kx$, where $K = \frac{2\pi}{a}$ with a is the period of modulation and $V_o$ is the amplitude of electric modulation.

To calculate the resistivity tensor $\rho_{\mu\nu}(\mu = x, \nu = y)$, we used Components of Conductivity tensor, which is given as:

$$\sigma = \begin{pmatrix} \sigma_{xx} & \sigma_{xy} \\ \sigma_{yx} & \sigma_{yy} \end{pmatrix} \quad (4.1)$$

The conductivity tensor is the sum of a diagonal and a non-diagonal elements. The diagonal elements cover the diffusive and collisional conductivity while the non-diagonal elements cover Hall conductivity. Using conductivity tensor and standard expression we calculated ([8] Vasilopoulos, 1989):

$$\rho_{xx} = \frac{\sigma_{yy}}{\sigma_{xx}\sigma_{yy} - \sigma_{xy}\sigma_{yx}}, \quad (4.2)$$

$$\rho_{yy} = \frac{\sigma_{xx}}{\sigma_{xx}\sigma_{yy} - \sigma_{xy}\sigma_{yx}} \quad (4.3)$$

$$\rho_{yx} = \frac{-\sigma_{yx}}{\sigma_{xx}\sigma_{yy} - \sigma_{xy}\sigma_{yx}} \quad (4.4)$$

Here we are interested only in non-diagonal elements, which give Hall conductivity. The 1D periodic potential is usually weak so Hall conductivity can be given as ([5] Krstajić, 2013):

$$\sigma_{yx} = \frac{2ihe^2}{\Omega} \sum_{\xi \neq \xi'} f_\xi (1 - f_{\xi'}) <\xi|v_x|\xi'><\xi|v_y|\xi'> \frac{1 - e^{\beta(E_\xi - E_{\xi'})}}{(E_\xi - E_{\xi'})^2} \quad (4.5)$$

Since

$f_\xi(1 - f_{\xi'})e^{\beta(E_\xi - E_{\xi'})} = f_{\xi'}(1 - f_\xi)$ and $\Omega \to A \equiv L_x L_y$, we arrive at following expression

$$\sigma_{yx} = \frac{2ihe^2}{A} \sum_{\xi \neq \xi'} (f_\xi - f_{\xi'}) \frac{<\xi|v_x|\xi'><\xi|v_y|\xi'>}{(E_\xi - E_{\xi'})^2} \quad (4.6)$$

In the presence of modulation the charge carrier acquire a mean velocity in x and y direction. Since $v_x = \frac{\partial H_o}{\partial p_x}$ and $v_y = \frac{\partial H_o}{\partial p_y}$

As we know that

$$H_o = v_f \, \sigma \cdot (\boldsymbol{p} + e\boldsymbol{A}) \quad (4.7)$$

$$H_o = v_f\{(\sigma_x p_x + \sigma_y p_y + \sigma_z p_z) + \sigma eA\} \quad (4.8)$$

The third component $\sigma_z p_z$ is zero, so we arrive at following results:

$$\boldsymbol{v_x} = v_f \sigma_x \quad (4.9)$$

$$v_y = v_f \sigma_y \quad (4.10)$$

Putting the value of $v_x$ and $v_y$ in equation (4.6) and solving the equation, hence

$$<\xi|v_x|\xi'> \quad = \quad <n, k_y|v_x|n'k_y> \tag{4.11}$$

$$<\xi|v_x|\xi'> \quad = \quad \int_0^{L_y}\int_{-\infty}^{+\infty} \psi^\dagger_{n,k_y} v_f \sigma_x \psi_{n',k_y} dxdy \tag{4.12}$$

$$<\xi|v_x|\xi'> \quad = \quad \frac{v_f}{2L_y l}\int_0^{L_y}\int_{-\infty}^{+\infty} \left(i\phi^*_{n-1}(\frac{x+x_o}{l})\phi^*_n(\frac{x+x_o}{l})\right)\begin{pmatrix}0 & 1\\ 1 & 0\end{pmatrix}\begin{pmatrix}i\phi_{n'-1}(\frac{x+x_o}{l})\\ \phi_{n'}(\frac{x+x_o}{l})\end{pmatrix} dxdy \tag{4.13}$$

$$<\xi|v_x|\xi'> \quad = \quad -\frac{v_f}{2l}\int_{-\infty}^{+\infty}\left(i\phi^*_{n-1}\left(\frac{x+x_o}{l}\right)\phi_{n'}\left(\frac{x+x_o}{l}\right) - i\phi^*_n\left(\frac{x+x_o}{l}\right)\phi_{n'-1}\left(\frac{x+x_o}{l}\right)\right)dxdy \tag{4.14}$$

Where $\phi_n = \frac{1}{\sqrt{2^n n!\sqrt{\pi}}} e^{-\frac{x^2}{2}} H_n(x)$ and $\phi_{n-1} = \frac{1}{\sqrt{2^{n-1}(n-1)!\sqrt{\pi}}} e^{-\frac{x^2}{2}} H_{n-1}(x)$ are harmonic

oscillator wave functions, n is the Landau level index.

By making substitution

$x' = \frac{x+x_0}{l}$ $\qquad dx = ldx'$

After substitution, $\phi_n$ and $\phi_{n-1}$ are harmonic oscillator wave functions becomes:

$$\phi_{n-1}(x') = \frac{1}{\sqrt{\sqrt{\pi}\, 2^{n-1}(n-1)!}} H_{n-1}(x')e^{-\frac{x'^2}{2}} \qquad \phi_n(x') = \frac{1}{\sqrt{\sqrt{\pi}\, 2^n n!}} H_n(x')e^{-\frac{x'^2}{2}}$$

$$\phi_{n-1}(x') = \frac{1}{\sqrt{\sqrt{\pi}\, 2^{n-1}(n-1)!}} H_{n-1}(x')e^{-\frac{x'^2}{2}}$$

Solving the equation (4.8), we arrive at the following mathematical Expression:

$$<\xi|v_x|\xi'> \quad = \quad <n, k_y|v_x|n'k_y>$$

$$< \xi | v_x | \xi' > \quad = \quad -\frac{iv_f}{2}(\delta_{n',n+1} - \delta_{n'n-1}) \tag{4.15}$$

Similarly we calculate $< \xi | v_y | \xi' >$

$$< \xi | v_y | \xi' > \quad = \quad < n, k_y | v_y | n' k_y >$$

$$= \int_0^{L_y} \int_{-\infty}^{+\infty} \psi^\dagger_{n.k_y} v_f \sigma_y \psi_{n',k_y} dxdy$$

$$< \xi | v_x | \xi' > = \frac{v_f}{2L_y l} \int_0^{L_y} \int_{-\infty}^{+\infty} \left( i\phi^*_{n-1}(\frac{x+x_o}{l}) \phi^*_n(\frac{x+x_o}{l}) \right) \begin{pmatrix} 0 & -i \\ i & 0 \end{pmatrix} \begin{pmatrix} i\phi_{n'-1}(\frac{x+x_o}{l}) \\ \phi_{n'}(\frac{x+x_o}{l}) \end{pmatrix} dxdy \tag{4.16}$$

$$< \xi | v_x | \xi' > \quad = \quad \frac{v_f}{2l} \int_{-\infty}^{+\infty} \left( \phi^*_{n-1}\left(\frac{x+x_o}{l}\right) \phi_{n'}\left(\frac{x+x_o}{l}\right) + \phi^*_n\left(\frac{x+x_o}{l}\right) \phi_{n'-1}\left(\frac{x+x_o}{l}\right) \right) dxdy \tag{4.17}$$

Solving the equation (4.0 8), we arrive at the following mathematical Expression:

$$< \xi | v_x | \xi' > \quad = \quad < n, k_y | v_x | n' k_y >$$

$$< \xi | v_x | \xi' > \quad = \quad \frac{v_f}{2}(\delta_{n',n+1} + \delta_{n'n-1}) \tag{4.18}$$

Substituting Eq (4.9) and (4.12) into Eq (4.6), the expression become:

$$\sigma_{yx} = \frac{2ihe^2}{A} \sum_{\xi \neq \xi'} (f_\xi - f_{\xi'}) \frac{-\frac{iv_f}{2}(\delta_{n',n+1} - \delta_{n'n-1})(\delta_{n',n+1} + \delta_{n'n-1})}{(E_\xi - E_{\xi'})^2} \tag{4.19}$$

$$\sigma_{yx} = \frac{v_f^2 \hbar e^2}{2L_y L_x} \sum_{n \neq n', k_y} \frac{(f_{n,k_y} - f_{n',k_y})}{(E_{n,k_y} - E_{n',k_y})^2} (\delta_{n',n+1} - \delta_{n'n-1}) \qquad (4.20)$$

$$\sigma_{yx} = \frac{v_f^2 \hbar e^2}{2L_y L_x} \sum_{n \neq n', k_y} \frac{(f_{n,k_y} - f_{n',k_y})\delta_{n',n+1} - (f_{n,k_y} - f_{n',k_y})\delta_{n'n-1}}{(E_{n,k_y} - E_{n',k_y})^2} \qquad (4.21)$$

Taking summation for $n' = n + 1$, in the first term and $n' = n - 1$ in the second term.

$$f_{n,k_y}\delta_{n',n+1} - f_{n',k_y}\delta_{n'n+1} = f_{n,k_y} - f_{n+1,k_y}$$

$$f_{n,k_y}\delta_{n',n-1} - f_{n',k_y}\delta_{n'n-1} = f_{n,k_y} - f_{n-1,k_y}$$

Hence

$$\sigma_{yx} = \frac{v_f^2 \hbar e^2}{2L_y L_x} \left\{ \sum_{n,k_y} \frac{(f_{n,k_y} - f_{n+1,k_y})}{(E_{n,k_y} - E_{n+1,k_y})^2} - \sum_{n,k_y} \frac{(f_{n,k_y} - f_{n-1,k_y})}{(E_{n,k_y} - E_{n-1,k_y})^2} \right\} \qquad (4.22)$$

In the second term of Eq (4.0 14) we change $n = n + 1$, we get

$$\sigma_{yx} = \frac{v_f^2 \hbar e^2}{2L_y L_x} \left\{ \sum_{n=0,k_y} \frac{(f_{n,k_y} - f_{n+1,k_y})}{(E_{n,k_y} - E_{n+1,k_y})^2} - \sum_{n=0,k_y} \frac{(f_{n+1,k_y} - f_{n,k_y})}{(E_{n,k_y} - E_{n+1,k_y})^2} \right\} \qquad (4.23)$$

$$\sigma_{yx} = \frac{v_f^2 \hbar e^2}{L_y L_x} \left\{ \sum_{n=0,k_y} \frac{(f_{n,k_y} - f_{n+1,k_y})}{(E_{n,k_y} - E_{n+1,k_y})^2} \right\} \qquad (4.24)$$

Since $E_\xi \equiv E_{n.k_y} = E_n + V_{n,B} \cos K x_o$ we obtain;

$$(E_{n,k_y} - E_{n+1,k_y})^2 = \hbar^2 \omega_g^2 \left[ \sqrt{n+1} - \sqrt{n} + \lambda_n, \cos K x_o \right]^2$$

Where

$$\lambda_n = \frac{V_e}{2\hbar \omega_g} e^{u^2/2} (L_{n+1}(u) - L_{n-1}(u))$$

Substituting these results in Eq (4.6) and expression becomes:

$$\sigma_{yx} = \frac{e^2 l^2}{ha} \sum_{n=0}^{\infty} \int_0^{L_y} \frac{f_{n,k_y} - f_{n+1,k_y}}{\left[\sqrt{n+1} - \sqrt{n} + \lambda_n, cosKx_o\right]^2} dk_y \qquad (4.25)$$

Using conductivity tensor and standard expression we calculated ([8] Vasilopoulos, 1989):

$$\rho_{yx} = \frac{-\sigma_{yx}}{\sigma_{xx}\sigma_{yy} - \sigma_{xy}\sigma_{yx}}$$

# Chapter Five

## Results and Discussion

The Eq (4.17) is the principal result of my work and this expression is for Hall conductivity for electrically modulated graphene system. The integral appear in Eq (4.17) are evaluated numerically and result are shown in Figure at temperature T= 2K for graphene system having density of electron $n_e = 3.0 \times 10^{15} m^{-2}$. The other parameter used in evaluating expression for Hall conductivity are shown in the table 4.20

To study the effect of one-dimensional periodic potential applied in the x direction of graphene system, we determine the correction to the Hall conductivity which is expressed as $\Lambda = \sigma_{xx}\sigma_{yy} - \sigma_{xy}\sigma_{yx} \approx \sigma_{yx}^2$ as shown in Figure 4.1 (b).

Figure 4.1 (b) show that the SdH oscillation are present in Hall conductivity. In addition Weiss oscillation are present but at higher magnetic field the SdH oscillation dominant.
To focus on the effect of temperature on Hall conductivity, we plotted the expression of Hall conductivity at two different temperatures: The solid curve shows the Hall conductivity for graphene as a function of magnetic field at T=2K and the dash curve shows the Hall conductivity at T=6K as shown in Figure 4.2 (a). Also correction to the Hall conductivity at two different temperature are shown in Figure 4.2 (b), which is clear indicted the presence of SdH oscillation.

Because SdH oscillations are strongly temperature as compare to Weiss oscillations. In addition at low magnetic field Weiss oscillation are also present in $\Delta\sigma_{yx}$ but effect of temperature on these oscillations is very small. So there are two factors which differentiate Weiss oscillations from SdH oscillations: First at low magnetic field the Weiss oscillations are dominant but as magnetic field is increased the SdH oscillations dominant and superimposed on Weiss oscillations. Second SdH are highly temperature dependent but Weiss oscillations have very effect of temperature. So at higher temperature the amplitude of SdH oscillations decreased significantly as compare to Weiss oscillations that's why we can easily differentiate both of them.

Table 1.1 : Parameters used to study Modulation in graphene system

| Parameter | Value |
|---|---|
| $V_e$ | 0.5meV |
| Period, a | 350nm |
| Fermi Energy, $E_F$ | 90.5mev |
| Density of Electron, $n_e$ | $3.0 \times 10^{15} m^{-2}$ |
| Impurity Density, $N_I$ | $2.5 \times 10^{11} m^{-2}$ |

Using Hall conductivity tensor we can find the elements of resistivity tensor. So the component of resistivity tensor $\rho_{yx}$ of graphene under one-dimensional periodic potential applied in the x direction as function of magnetic field and also correction to Hall resistivity as shown in Figure (4.3) .Furthermore, to study the effect of temperature, resistivity and correction to resistivity at two different temperatures as shown in Figure 4.4: The solid curve shows the resistivity for graphene as a function of magnetic field at T=2K and the dash curve shows the resistivity at T=6K.

## 5.1 Integral Quantum Hall Effect

From the Figure 4.1 to 4.4 we seen that Hall Conductivity and resistivity is not affected by modulation. If we take derivative of $\rho_{yx}$ and plot as a function of magnetic field, one might infer that Hall conductivity and resistivity carries modulation effects.

For unmodulated graphene, U(x) =0

$$\lambda_n = \frac{V_e}{2\hbar\omega_g} e^{u^2/2}(L_{n+1}(u) - L_{n-1}(u)) = 0$$

On solving Equation (4. 1) become

$$\sigma_{yx} = 4\frac{e^2}{h}(N+1)$$

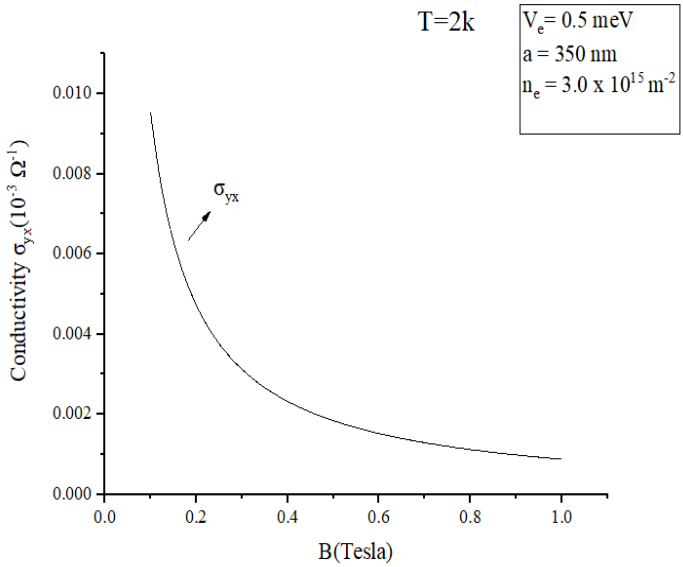

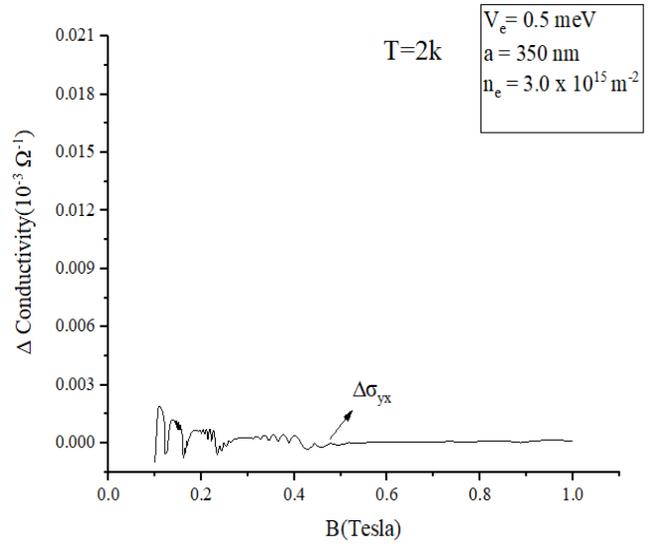

Figure 4.2(a): The conductivity ($\sigma_{yx}$) for electrically modulated graphene as a function of magnetic field at Temperature (T=2K).

Figure 4.1(b): The change in the conductivity ($\sigma_{yx}$) for electrically modulated graphene as a function of magnetic field at Temperature (T=2K).

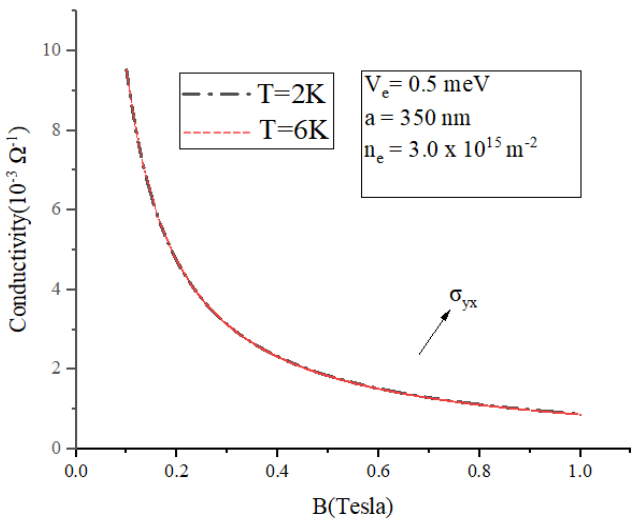

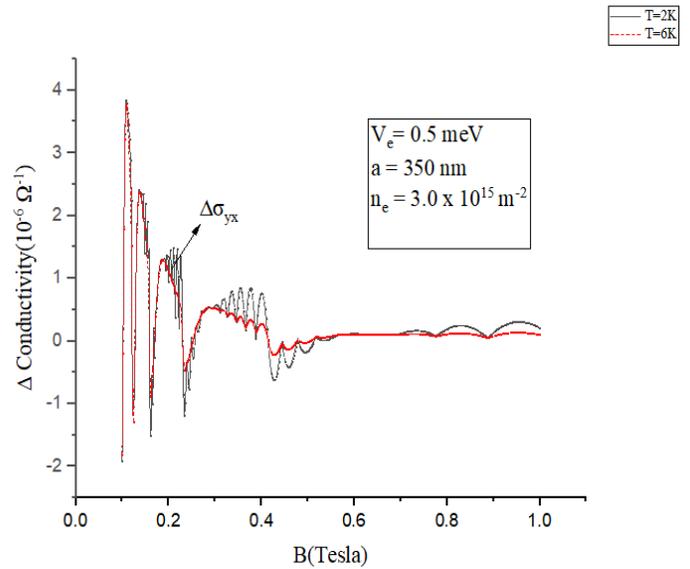

Figure 4.3(a): the conductivity ($\sigma_{yx}$) for electrically modulated graphene as a function of magnetic field at two different Temperature (T=2K: solid curve and T=6K: dash curve).

Figure 4.4(b): The conductivity ($\sigma_{yx}$) for electrically modulated graphene as a function of magnetic field at two different Temperature (T=2K: solid curve and T=6K: dash curve).

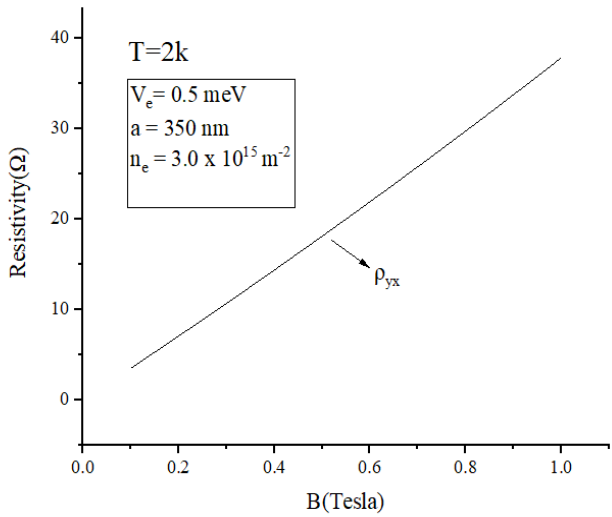

Figure 4. 6 (a): The resistivity ($\rho_{yx}$) for electrically modulated graphene as a function of magnetic field at Temperature (T=2K).

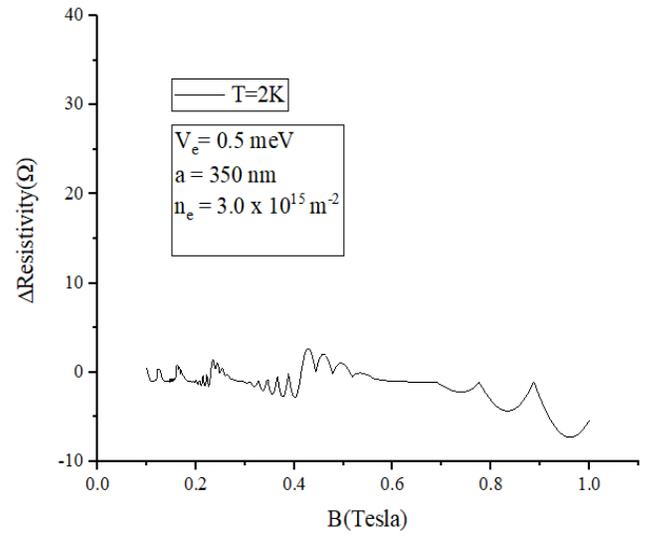

Figure 4.5 (b): The resistivity ($\rho_{yx}$) for electrically modulated graphene as a function of magnetic field at Temperature (T=2K).

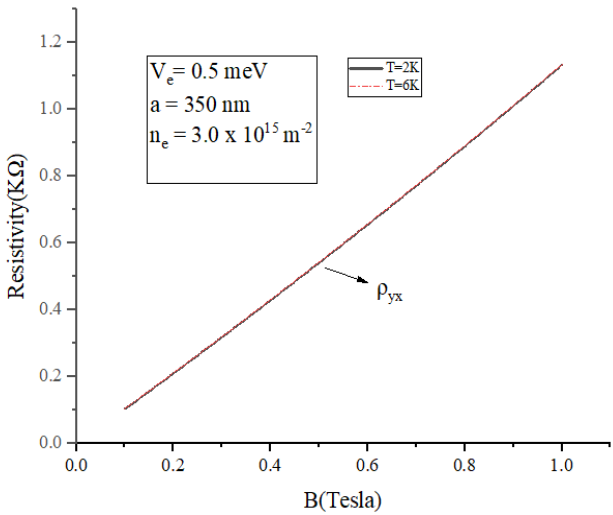

Figure 4.8(a): The resistivity ($\rho_{yx}$) for electrically modulated graphene as a function of magnetic field at Temperature (T=2K: solid curve and T=6K: dash curve).

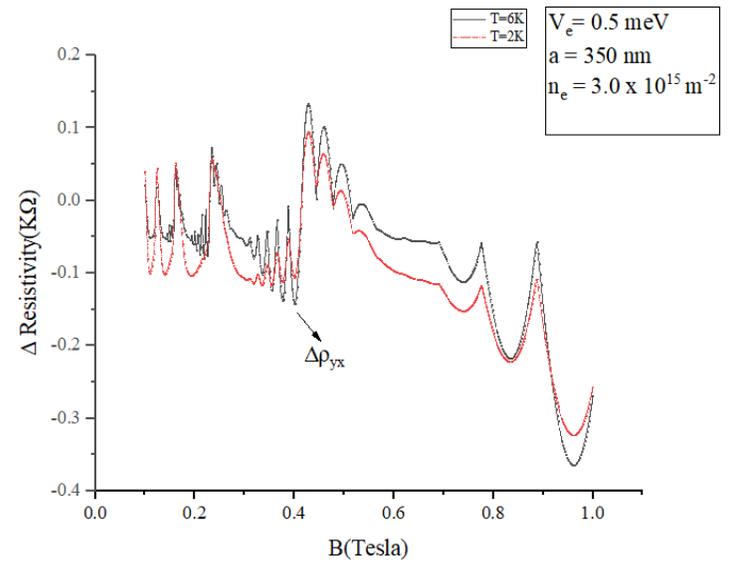

Figure 4.7(b): The change in resistivity ($\rho_{yx}$) for electrically modulated graphene as a function of magnetic field at Temperature (T=2K: solid curve and T=6K: dash curve).

However, in the presence of modulation, the term $[\sqrt{n+1} - \sqrt{n} + \lambda_n cosKx_o]$ in Eq (4.017) gives energy difference between two Landau levels which oscillate with magnetic field and leads to oscillation in $\sigma_{yx}$ and $\rho_{yx}$. As modulation is very weak, so oscillations appear in $\sigma_{yx}$ and $\rho_{yx}$ due the term $\lambda_n cosKx_o$ is very small. If we take the derivative of $\rho_{yx}$ with respect to magnetic field these oscillations become more prominent and finally lead to quantum Hall effect as shown in Figure 4.9

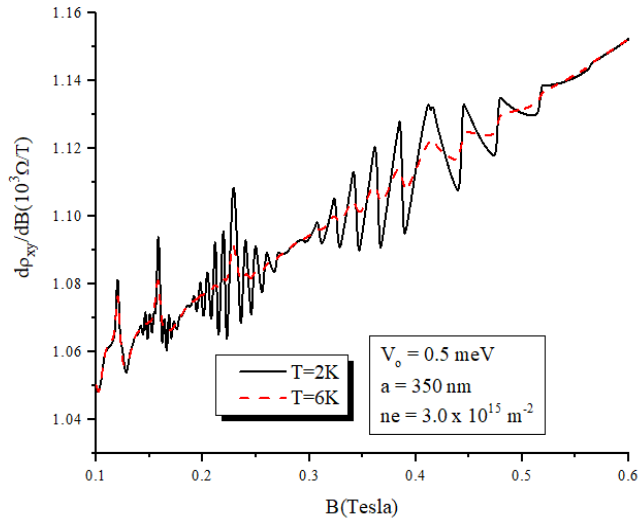

Figure 4.9: The derivative (d $\rho_{yx}$/dB) for electrically modulated graphene as a function of magnetic field at two different Temperature (T=2K: solid curve and T=6K: dash curve).

## 5.2 Conclusion

In this work, we have investigated the effect on Hall conductivity due to weak electrically modulated graphene under perpendicular magnetic field B. We have carried out a detailed study of modulation effect on Dirac electrons, which determine its electrical transport properties. In a result of electric modulation Commensurability (Weiss) oscillations and Shubnikov de Haas (SDH) oscillations are prevailed. Due to these oscillations integer quantum Hall is observed.